\documentclass[12pt]{article}

\newcommand{\bbr}{I\!\! R}

\newcommand{\x}{arXiv:}
\newcommand{\m}{\mathrm}
\newcommand{\be}{\begin{equation}}
\newcommand{\ee}{\end{equation}}
\newcommand{\ba}{\begin{eqnarray}}
\newcommand{\ea}{\end{eqnarray}}

\usepackage{graphicx}
\usepackage{amssymb}
\usepackage{amsmath}
\usepackage[T1]{fontenc} 
\usepackage[ansinew]{inputenc} 
\usepackage[nosort]{cite}
\newcommand{\inbar}{\vrule height1.57ex width.4pt depth0pt}
\newcommand{\SW}{\relax{\hbox{$\ \inbar\kern-.285em{\rm S}$}}}

\begin{document}
\thispagestyle{empty}
\begin{center}

\null \vskip-1truecm \vskip2truecm

{\Large{\bf \textsf{Extremal Instability for Topological Black Holes}}}

{\large{\bf \textsf{}}}

{\large{\bf \textsf{}}}

\vskip1truecm

{\large \textsf{Brett McInnes}}

\vskip1truecm

\textsf{\\  National
  University of Singapore}

\textsf{email: matmcinn@nus.edu.sg}\\

\end{center}
\vskip1truecm \centerline{\textsf{ABSTRACT}} \baselineskip=15pt
\medskip

The initial idea underlying the Weak Gravity Conjecture is that extremal black holes must always be ``unstable'', in the sense that they should slowly decay by emitting either particles or smaller black holes. Here we show that, when this idea is applied to the \emph{planar} asymptotically AdS black holes which play a central role in applications of holography, the result, via gauge-gravity duality, is a prediction that there should exist a lower bound on the possible densities of cold strongly coupled matter. Recent observations of neutron stars suggest that, in many cases, even the extreme densities in their cores may not be sufficient to generate quark matter, showing that there is indeed a (very high) lower bound on the possible density of cold quark matter.

\newpage

\newpage

\addtocounter{section}{1}
\section* {\large{\textsf{1. The Density of Cold Strongly Coupled Matter}}}
Quark matter \cite{kn:conf} can exist under two very different sets of circumstances: either at extreme temperatures, as in the plasmas generated by heavy-ion collisions, or at extreme \emph{densities}, as in (perhaps) the cores of very massive neutron stars.

From the beginning \cite{kn:urwitten}, it has been hoped that the AdS/CFT duality \cite{kn:casa,kn:nat,kn:bag}, applied to asymptotically locally AdS black hole spacetimes, might throw some light on deconfinement due to high temperatures. For AdS-Schwarzschild black holes, there is indeed a lower bound on the range of possible Hawking temperatures, below which there is a phase transition. This means that the dual system, a conformal field theory modelling a quark-gluon plasma, cannot exist at low temperatures, reflecting confinement (in the model, if not in the actual plasma). Holographic models of thermally-induced deconfinement continue to be a field of active investigation: for recent examples, see \cite{kn:nickevans,kn:rocha}.

Deconfinement due to high densities is considerably less well-understood. For example, it can be argued that recent observations of neutron star radii \cite{kn:noquark} indicate that, contrary to many theoretical expectations, the densities found in the cores of neutron stars observed hitherto, on the order of 10$^{15}$ g/cm$^3$, may \emph{not} be sufficient to form quark matter (though this remains an open question for the most massive neutron stars \cite{kn:robmann}). Evidently there is a pressing need for a better understanding of the \emph{unexpectedly high lower bound on the density} of (cold) quark matter that these observations prove to exist.

Very sophisticated holographic models of cold quark matter have been developed (see for example \cite{kn:nordita,kn:matti,kn:niko} and their references), but it is far from obvious how one might use these models to account for this lower bound. The problem can be formulated in elementary terms as follows: in the simplest models, where the bulk geometry is just that of a five-dimensional AdS-Reissner-Nordstr\"{o}m black hole, the energy density (or more precisely, the enthalpy density\footnote{It is often argued \cite{kn:sourya} that the mass of an asymptotically locally AdS spacetime corresponds to the black hole enthalpy. Although this does not affect our discussion materially, we will accept this interpretation here, because enthalpy in this case is naturally related to the (quark) chemical potential which, in the quark matter literature, is often used to describe ``density''.}) of the field theory on the conformal boundary is dual to the mass per unit horizon volume of the bulk black hole, and there seems to be no reason for this latter quantity to be bounded below. We must ask: what physical effect (in the bulk) could possibly enforce such a bound?

We wish to propose that the relevant effect is \emph{Extremal Instability}: that is, the conjectured process which requires that all black holes sufficiently close\footnote{Note that we do not require the black hole to be exactly extremal; it may well be that such objects do not actually exist (see for example \cite{kn:wilczek,kn:naresh,kn:haoong}); we regard them here as a useful approximation to the physical, near-extremal case.} to extremality are necessarily unstable to some effect which causes them to emit particles, branes, or small black holes. This claim can be motivated by observations: there are many observed black holes which are very close to being extremal (for example, it has been argued \cite{kn:cygnus} that Cygnus X-1 contains a black hole with dimensionless spin parameter at least $0.9696,$ where unity represents extremality), but no black hole-like object has been observed with dimensionless spin parameter beyond unity. This may of course be a straightforward astrophysical effect (see for example \cite{kn:benson}), but it is also possible that sufficiently near-extremal black holes are unstable for fundamental physical reasons. We will assume this here, and that it applies to \emph{all} near-extremal black holes, whether extremality be due to angular momentum or to charge.

This concept of Extremal Instability came to prominence in connection with the \emph{Weak Gravity Conjecture} \cite{kn:motl,kn:palti,kn:rude} (for the asymptotically locally AdS context relevant here, see \cite{kn:qing,kn:naka1,kn:mig,kn:crem,kn:agar,kn:naka2,kn:102,kn:103,kn:104,kn:105}). In fact, the WGC was originally motivated in this way, though doubts regarding this motivation have been expressed (see for example \cite{kn:heid}), and the question continues to be debated.

However, the Extremal Instability conjecture is actually independent of the WGC; for example, it can be applied \cite{kn:104,kn:105} to cases (like extremal Kerr and extremal AdS-Kerr) with no interpretation in terms of the ``weakness'' of gravity. In this work, we focus on the Extremal Instability conjecture (which is of considerable independent interest), and we do not focus on the WGC itself, except tangentially as an application of Extremal Instability.

The idea, then, is that if we use cold (that is, near-extremal) black holes in a holographic model of \emph{cold} quark matter, then their conjectured instability might impose interesting conditions on the dual system, possibly including a lower bound on the enthalpy density.

When the event horizon is spherical, the instability of extremal asymptotically flat black holes has the familiar (from studies of the WGC) consequence that the mass of the emitted object, in Planck units, is smaller than (a fixed multiple of) its charge: hence ``weak'' gravity. The conclusion is precisely the same in the spherically symmetric, asymptotically AdS case \cite{kn:102}, essentially because the bound in that case is imposed by a limit in which the two local geometries tend to coincide.

As is well known, however, a locally asymptotically AdS black hole need not have a spherical event horizon \cite{kn:lemmo}: in particular, the event horizon can be \emph{flat}. This case\footnote{Usually one says that the \emph{event horizon} is flat. However, in fact the spatial sections transverse to the radial direction (the ``$r = $ constant'' spatial sections) have this geometry and topology throughout the entire spacetime, and the conformal boundary is similarly affected. Thus we prefer to speak of \emph{flat transverse spatial sections}.} is in fact \emph{the one of primary interest in the holographic context}, since the spacetime at infinity (in the case of a five-dimensional bulk) in this case then has the topology of $\bbr^4$ (perhaps with spatial sections compactified), and this is of course the correct topology for most applications.

It is very important to note that these AdS black holes have \emph{no} asymptotically flat counterparts: they belong to an entirely distinct, indeed \emph{disjoint}, class of solutions of the Einstein equations. While they are often regarded as a ``scaling limit'' of spherical AdS black holes, it turns out that certain quantities \emph{do not behave continuously under this limit.} We will see that one such quantity plays a central role in the analysis of the consequences of extremal instability, which therefore takes an entirely different form in the planar/toroidal case to the form it takes in the more familiar asymptotically flat, spherical context.

We find that the conjectured instability of extremal planar black holes has radically different consequences to those arising in the spherically symmetric case. Instead of imposing an \emph{upper} bound on the mass/charge ratio of the emitted object, the resulting inequality is best interpreted in this case in the opposite way: the minimal possible mass/charge ratio of the emitted object puts a \emph{lower} bound on the charge parameter of an extremal black hole of this sort. This translates easily to a lower bound on the enthalpy density of the dual theory, just as we hoped.

We stress that this only establishes the \emph{existence} of such a bound: we are far from being able to evaluate the bound explicitly, so as to be able to determine whether it might be useful in explaining the lower bounds deducible from data on neutron stars. Furthermore, the bound is derived in the most basic, and consequently unrealistic, holographic models. But the arguments leading to this bound are so general and simple that it is reasonable to hope that they can be carried over to the much more elaborate and realistic models discussed in \cite{kn:nordita,kn:matti,kn:niko}, if the Extremal Instability principle can be incorporated into them.

We begin with a brief review of locally AdS$_5$-Reissner-Nordstr\"om black holes with flat transverse spatial sections.

\addtocounter{section}{1}
\section* {\large{\textsf{2. AdS$_5$-Reissner-Nordstr\"om Black Holes With Flat Transverse Spatial sections}}}
The ``topological'' locally asymptotically AdS$_5$-Reissner-Nordstr\"om metric with flat (toroidal or planar) transverse spatial sections takes the form (with the zero superscript indicating that the transverse spatial sections are indeed flat):
\begin{flalign}\label{A}
g(\m{AdS_5RN^0})\;=\;&-\,\left({r^2\over L^2}\,-\,{16\pi M^*\ell_5^3\over 3r^2}\,+\,{4\pi k_5 Q^{*2}\ell_5^3\over 3r^4}\right)\m{d}t^2\,+{\m{d}r^2\over {r^2\over L^2}\,-\,{16\pi M^*\ell_5^3\over 3r^2}\,+\,{4\pi k_5 Q^{*2}\ell_5^3\over 3r^4}}\\ \notag \,\,\,\,&\,+\,r^2\left(\m{d}\theta^2 \,+\,\m{d}\phi^2\,+\,\m{d}\psi^2\right).
\end{flalign}
Here $k_5$ is the five-dimensional\footnote{Notice that, unlike its four-dimensional counterpart, $k_5$ has units of length (in units such that charge and entropy are dimensionless, and mass has units of inverse length). It is an independent parameter, which must not be ignored by being ``set equal to unity''. See also \cite{kn:myers}. We will see later that $k_5$ has a definite holographic interpretation.} Coulomb constant, $L$ is the AdS curvature length scale, $\ell_5$ is the gravitational length scale for AdS$_5$, and the coordinates $(\theta, \phi, \psi)$ are circular, with periodicity $2\pi K$; here $K$ is dimensionless and can take any positive value; the planar case is obtained in the limit $K \rightarrow \infty$. Thus, in the toroidal case, the three-dimensional volume of any transverse spatial section is given by $Wr^3$, where $W \equiv \left(2\pi K\right)^3$.

The electric field in this spacetime is
\begin{equation}\label{B}
E \;=\;{k_5Q^*\over r^3}.
\end{equation}

The ``mass and charge parameters'' $M^*$ and $Q^*$ are related but \emph{not} equal to the mass and charge of the black hole (though they do have the units of mass and charge respectively). In fact, the case of principal interest to us here is the one with planar transverse spatial sections, for which the mass and charge are formally infinite, so the interpretations of $M^*$ and $Q^*$ are particularly in need of clarification in that case.

This is explained in a general way in \cite{kn:peldan} by means of a deep study of the Hamiltonian at infinity, but in this specific case it can be explained in a very elementary manner; this is worth reviewing, because the Extremal Instability hypothesis itself plays the simplifying role.

We begin with the toroidal case, returning to the planar case below. Since, as above, the volume of the transverse spatial sections is $Wr^3$, Gauss' law applied to equation (\ref{B}) implies that the physical charge of the black hole, $Q$, is given as
\begin{equation}\label{C}
Q \;=\; EWr^3/k_5 = WQ^*.
\end{equation}
Thus we see that $Q^* = Q/W.$

In order to find the mass of the black hole, we need its Hawking temperature, given (from (\ref{A})) by
\begin{equation}\label{D}
T\;=\;{1\over \pi}\left({r_{\textsf{H}}\over L^2}\;-\;{2\pi k_5Q^{* 2}\ell_5^3 \over 3 r_{\textsf{H}}^5}\right),
\end{equation}
where the event horizon is located at $r = r_{\textsf{H}}$, and where we have eliminated $M^*$ by using the definition of the event horizon:
\begin{equation}\label{E}
{r_{\textsf{H}}^2\over L^2}\,-\,{16\pi M^*\ell_5^3\over 3r_{\textsf{H}}^2}\,+\,{4\pi k_5 Q^{*2}\ell_5^3\over 3r_{\textsf{H}}^4} = 0.
\end{equation}

We also need the entropy of the black hole, given (again in the toroidal case) by
\begin{equation}\label{F}
S\;=\;{Wr_{\textsf{H}}^3 \over 4 \ell_5^3}.
\end{equation}
The mass will be found by integrating the usual thermodynamic relation $dM = TdS:$
\begin{equation}\label{G}
dM\;=\;{3W\over 4\pi \ell_5^3}\left({r_{\textsf{H}}^3\over L^2}\;-\;{2\pi k_5Q^{* 2}\ell_5^3 \over 3 r_{\textsf{H}}^3}\right)dr_{\textsf{H}},
\end{equation}
so we have
\begin{equation}\label{H}
M\;=\;{3W\over 16\pi \ell_5^3}\left({r_{\textsf{H}}^4\over L^2}\;+\;{4\pi k_5Q^{* 2}\ell_5^3 \over 3 r_{\textsf{H}}^2}\right)\;+\;C,
\end{equation}
where $C$ is a constant of integration.

By setting $T = 0$ for the extremal case, we see (from equation (\ref{D})) that the extremal value of $r_{\textsf{H}},$ which we name $r_{\textsf{H}}^{\textsf{ext}},$ has a simple form:
\begin{equation}\label{I}
r_{\textsf{H}}^{\textsf{ext}}\;=\;\left({2\pi k_5\over 3}\right)^{1/6}\,\sqrt{\ell_5}\,L^{{1\over 3}}Q^{*{1\over 3}}.
\end{equation}
Substituting this into (\ref{H}), we see that the mass in the extremal case is equal to a certain multiple of $Q^{*{4\over 3}}$, plus $C$. Since we know that $Q^*$ is a multiple of the charge, this means that $C$ is the mass of an electrically neutral black hole with zero temperature. If such an object could exist, it would be absolutely stable, contradicting the principle of Extremal Instability, and so we conclude that $C = 0$. Thus we have
\begin{equation}\label{J}
M\;=\;{3W\over 16\pi \ell_5^3}\left({r_{\textsf{H}}^4\over L^2}\;+\;{4\pi k_5Q^{* 2}\ell_5^3 \over 3 r_{\textsf{H}}^2}\right).
\end{equation}
Using (\ref{E}), one sees that the right side of this equation is just $WM^*;$ so we have simply $M^* = M/W.$ (It turns out that the same result holds in the spherical case, but with $W = 2\pi^2,$ the volume of the unit three-sphere.)

The direct physical interpretations of $M^*$ and $Q^*$ are obtained by noticing that the mass of the black hole per unit horizon volume is just $M/\left(Wr_{\textsf{H}}^3\right) = M^*/r_{\textsf{H}}^3,$ and similarly the charge of the black hole per unit horizon volume is $Q^*/r_{\textsf{H}}^3.$ In this way, we can avoid referring to the actual mass and charge. (Notice that the event horizon ``radius'' $r_{\textsf{H}}$ is determined by the values of $M^*$ and $Q^*,$ not $M$ and $Q$.)

This construction is very useful, because it allows us to let $K \rightarrow \infty$ while simultaneously taking the mass and charge to infinity, maintaining \emph{finite} values for $M^*$ and $Q^*$. In this way we can also allow for planar, as well as toroidal, transverse spatial sections: in particular, the planar black hole has a well-defined mass per unit horizon volume, $M^*/r_{\textsf{H}}^3$. For this reason, we state our main results in terms of $M^*$ and $Q^*$.

We now come to the condition for Cosmic Censorship to hold. This is superficially an elementary question, but there is a crucial subtlety here.

It will be useful to treat the spherical and toroidal cases in a unified way. To that end, in the spherical case let us define $M^* = M/(2\pi^2)$, $Q^* = Q/(2\pi^2)$. (As mentioned above, $2\pi^2,$ the volume of the unit three-sphere, plays the role of $W$ here.) Then the equation defining the value of the radial coordinate at the event horizon, $r_{\textsf{H}},$ can be written as
\begin{equation}\label{KK}
{x^3\over L^2}\;+\;px^2\;-\;AM^*x\;+\;BQ^{*2}\;=\;0,
\end{equation}
where $x = r_{\textsf{H}}^2,$ where $p = 0, 1$ in the toroidal and spherical cases respectively, and where $A = 16\pi \ell_5^3/3,\; B = 4\pi k_5\ell_5^3/3.$ (See (\ref{E}) above for the toroidal case.)

The discriminant \cite{kn:disc} of the general cubic $ax^3 + bx^2 + cx + d$ is given by
\begin{equation}\label{KKK}
\mathcal{D} \;\equiv\; 18abcd - 4b^3d + b^2c^2 - 4ac^3 - 27a^2d^2.
\end{equation}
For the cubic in (\ref{KK}), this is
\begin{equation}\label{KKKK}
\mathcal{D}\;=\;-\;4pBQ^{*2}\;+\;pA^2M^{*2}\;-\,{18pAB\over L^2}M^*Q^{*2}\;\;+\;{4A^3\over L^2}M^{*3}\;-\;{27B^2\over L^4}Q^{*4}.
\end{equation}
Cosmic Censorship is now expressed, in both cases, as $\mathcal{D} \geq 0.$

When $p = 1,$  we can take the limit as $L \rightarrow \infty$, and then we see that extremal black holes have a charge parameter which is a fixed multiple of the mass parameter: this is the familiar result in the asymptotically flat case. We obtain the same result when $M^*$ and $Q^*$ are small, because the third, fourth, and fifth terms on the right are either cubic or quartic and can be neglected in comparison with the first and second terms. That is, the spherical AdS case goes over smoothly to the asymptotically flat case in both limits.

The $p = 0$ case is very different. Here, if we attempt to take the limit $L \rightarrow \infty$ in equation (\ref{A}), we find that the ``time'' coordinate $t$ is spacelike at \emph{large} values of $r$, and that the metric is not asymptotically flat. Thus, it does not make sense to take this limit in this case: toroidal/planar AdS black holes have \emph{no} asymptotically flat analogues.

In the toroidal/planar case, the first three terms (\ref{KKKK}) are exactly zero, and so \emph{all connection with the AdS spherical case is lost}. We see at once that $M^*$ and $Q^*$ are no longer proportional to each other for extremal black holes of this kind, whether $M^*$ and $Q^*$ be large or small; instead, the mass parameter is bounded by a multiple of the 4/3 \emph{power} of the charge parameter. As far as the criterion for Cosmic Censorship is concerned, there is simply no limit in which spherical and planar/toroidal AdS black holes go over to each other in a \emph{continuous} manner.

The precise condition for Cosmic Censorship to hold for toroidal/planar AdS$_5$-Reissner-Nordstr\"om black holes is
\begin{equation}\label{L}
M^*\;\geq \;{3\over 16\,\ell_5}\left({12\,k_5^2\,Q^{*4}\over \pi L^2}\right)^{{1\over 3}}.
\end{equation}
We see in particular that the ratio $M^*/Q^*$ is \emph{not} bounded below by a constant (at least, not by classical Censorship alone).

Black holes with flat transverse spatial sections simply do not exist in the asymptotically flat case, so there is no occasion for surprise that AdS black holes with flat transverse spatial sections cannot be understood by extrapolating our experience with asymptotically flat black holes. It is more surprising however that, as far as the condition for Cosmic Censorship is concerned, AdS black holes with flat transverse spatial sections \emph{also} cannot be fully understood by extrapolating from the \emph{AdS} spherical case. This second point has the drastic consequences we are about to discuss.

\addtocounter{section}{1}
\section* {\large{\textsf{3. Extremal Instability for AdS$_5$-Reissner-Nordstr\"om Black Holes With Flat Transverse Spatial sections}}}
Consider a positively charged extremal AdS$_5$-RN black hole with flat transverse spatial sections: its mass and charge parameters saturate the inequality (\ref{L}). We will temporarily assume that the transverse spatial sections have been compactified to a torus, but this is a mere technicality: see below\footnote{Like their spherical counterparts, locally asymptotically AdS black holes with flat, toroidal transverse spatial sections have a thermal phase transition \cite{kn:surya}; but the transition occurs at a temperature which can be made arbitrarily small by adjusting the spatial compactification scale to be sufficiently large. Since we do not wish to have such a transition in this application, and since we do not need the spatial sections at infinity to be compact, we do in fact want to take $K \rightarrow \infty$. Thus the planar case is stable with respect to temperature, and we can use it to investigate stability with respect to density.}.

We now wish to consider the possibility that near-extremal AdS$_5$-RN black holes with flat transverse spatial sections (which we approximate by the exactly extremal case) are not stable. Some caution is called for here, because, as is well known, \emph{some} AdS black holes are, if reflective boundary conditions are imposed at infinity, able to come into equilibrium with their own Hawking radiation, and it is possible that an analogous phenomenon could occur here. This has led to some doubts as to the meaning of black hole ``instability'' in the AdS case.

We will argue that these doubts are misplaced. Consider first the case of Hawking radiation.

Note first that, even in the case of reflective boundary conditions, some AdS black holes evaporate completely before their Hawking radiation is able to return \cite{kn:ruong}. Observe in that connection that the disintegration of any black hole well be associated with thermodynamically irreversible processes in the interior, along the lines explored in detail in \cite{kn:empa}. Thus it is not always the case that AdS black holes settle down to a static state when Hawking radiation is taken into account.

Second, if we wish, we can replace reflective boundary conditions by adjoining an asymptotically flat spacetime at infinity, as has recently been done very effectively in discussions of the information paradox \cite{kn:pen,kn:alm1,kn:alm2}. In short, there are well-established ways in which the decay of AdS black holes can be defined unambiguously.

A mechanism of AdS black hole decay more directly relevant here was studied by Seiberg and Witten \cite{kn:seiberg} (see \cite{kn:niko1,kn:niko2,kn:oscar} for more recent studies). Here one finds that, under certain conditions which in this case amount \cite{kn:107} to nearness to extremality, that the action of certain branes in asymptotically AdS spacetimes can be negative in the asymptotic region, so that they can be created and moved away from the black hole, towards infinity \cite{kn:maldacena}. The great difference from the Hawking radiation case is that this mechanism can work for arbitrarily \emph{cold} black holes, and is therefore relevant to the near-extremal case.

Again, the point is that one has a definite (if not yet fully understood) scenario for near-extremal AdS black hole decay. Henceforth we proceed on the assumptions that Extremal Instability is well-defined for AdS black holes, and that it is valid; though we do not claim that the underlying mechanism is as well-understood as Hawking radiation.

Although the WGC is not our primary concern here, we digress to it briefly in order to discuss a question which has given rise to some confusion.

Some authors (see most notably \cite{kn:naka1}) have attempted to extend the WGC to the asymptotically AdS case by guessing a generalization of the asymptotically flat WGC condition, which in five dimensions states
\begin{equation}\label{LL}
{m\ell_5\over q} \;\leq\; {1\over 4}\,\sqrt{{3k_5\over \pi \ell_5}},
\end{equation}
where $m$ and $q$ are the mass and charge of some distinguished particle in the bulk. The suggestion in \cite{kn:naka1} is that one should simply replace $m$ with $\Delta/L$, where $\Delta$ is the scaling dimension of the boundary operator dual to the field associated with this particle; the reasoning is that this is the natural quantity to consider from the AdS/CFT perspective. Thus one obtains
\begin{equation}\label{LLL}
{\Delta\ell_5\over qL} \;\leq\; {1\over 4}\,\sqrt{{3k_5\over \pi \ell_5}}
\end{equation}
as the proposed statement of the WGC in the AdS case.

The justification for this is that $\Delta/L$ goes over to $m$ when $L \rightarrow \infty$: it is given in the case of (for example) scalar fields by
\begin{equation}\label{LLLL}
{\Delta\over L}\;=\;{2\over L}\,+\,\sqrt{{4\over L^2}\,+\,m^2}.
\end{equation}
Notice however that this formula varies according to the spin. Thus, by proceeding in this way, we will obtain a whole \emph{collection} of WGC bounds, depending on the spin of the distinguished particle, which remains to be determined.

We have three further comments to make on this proposal.

First, as was recently stressed in \cite{kn:heid}, there are infinitely many other possible formulae, even for particles of given spin, which reduce to the asymptotically flat version of the WGC when $L \rightarrow \infty$, and it has not been explained why this one should be favoured. The authors of \cite{kn:heid} in fact dismiss the proposal on these grounds.

Secondly, $\Delta/L$ is subject to universal inequalities: for example, again in the case of scalar fields, one has
\begin{equation}\label{LLLLL}
{\Delta\over L}\;\geq\;{2\over L}.
\end{equation}
This is just the Breitenlohner-Freedman bound. (See \cite{kn:wit}; a similar bound applies to a vector field, with $1/L$ on the right.) Substituting this into (\ref{LLL}), we obtain
\begin{equation}\label{LLLLLL}
q^2\;\geq {64\pi\ell_5^3\over 3k_5L^2}.
\end{equation}
That is, the proposal implies postulating the existence of \emph{independent} fundamental lower bounds on the ``mass'' (inequality (\ref{LLLLL})) and the charge, depending on the asymptotic curvature scale $L$ and on the five-dimensional Newton constant $\ell_5^3$ and the five-dimensional Coulomb constant $k_5.$ No physical justification for this procedure (or identification of the constant on the right side of (\ref{LLLLLL})) suggests itself, and this is far from being a straightforward generalization of the asymptotically flat version of the WGC.

Finally, none of this will begin to work in the case of interest to us here, toroidal/planar black holes. For, as we have stressed repeatedly, one does \emph{not} expect ``reduction to the asymptotically flat case when $L \rightarrow \infty$'' to be of any use as a guiding principle in this case.

In view of all this, we find it far more natural and physical simply to follow the procedure used in the original work on the WGC, \cite{kn:motl} and \cite{kn:kats}. This is a straightforward implementation of Extremal Instability: the black hole is then compelled to emit an object which is moving relative to it (presumably at relativistic speed), so that the mass of the black hole decreases by a larger amount than the mass of the emitted object (which is as well-defined in an AdS spacetime as in any other). These extremely simple assumptions, together with charge conservation, are all we need to determine the consequences of Extremal Instability in this case. It is true that, in the AdS/CFT context, it is natural to formulate the \emph{results} in terms of $\Delta$ (using formulae such as (\ref{LLLL})), and indeed we will do this later (see the discussion at the end of Section 4, below).

Let us proceed. The emitted object ---$\,$ which could be a particle, a brane, or \cite{kn:kats, kn:NAH} a black hole\footnote{This black hole presumably respects the topology of the ambient spacetime; that is, in the present discussion, it too has a toroidal event horizon, with the same compactification parameter, $K$.}---$\,$ of mass $m$ and (without loss of generality) positive charge $q$. Subsequently the mass and charge of the original black hole become $(M + \delta M, Q + \delta Q)$, where $\delta M$ and $\delta Q$ are of course negative. Then, by the triangle inequality, we have
\begin{equation}\label{M}
m\; < \;-\delta M.
\end{equation}
Charge conservation, $q = - \delta Q,$ gives us
\begin{equation}\label{N}
{m \over q}\, < \,{\delta M \over \delta Q} \;=\; {\delta M^* \over \delta Q^*};
\end{equation}
we can now drop (if we wish) the assumption that the transverse spatial sections have been compactified. (If the emitted object is a black hole with parameters $m^*$ and $q^*,$ then the left side of (\ref{N}) should be written as $m^*/q^*.$)

As usual in discussions of the WGC, we assume that the original black hole satisfied and continues to satisfy Cosmic Censorship after emitting the object: that is, the point in the $(Q^*, M^*)$ plane representing the original black hole moves in such a way as to stay on or above the curve (see the right side of (\ref{L}), above) describing extremality. Regarding extremal values of $M^*$ as defining a function of $Q^*,$ then we must have
\begin{equation}\label{O}
{\delta M^* \over \delta Q^*} \;\leq \; {\m{d}M^* \over \m{d} Q^*};
\end{equation}
combining this with (\ref{N}), and differentiating the right side of (\ref{L}) ---$\,$ notice that we are using classical Cosmic Censorship at this point ---$\,$ we have
\begin{equation}\label{P}
{m\ell_5\over q}\; < \;{1\over 4}\left({12\,k_5^2\,Q^*\over \pi L^2}\right)^{{1\over 3}}.
\end{equation}
This has been expressed in such a manner that both sides are dimensionless.

\emph{From this point onwards we confine attention to the case where the emitted object is not a black hole.} The case in which it is a black hole will be discussed elsewhere; it differs in important ways from the particle/brane case.

The left side of (\ref{P}) is ambiguous as it stands, since many varieties of brane/particles could be emitted, and (more seriously) since the emitted brane/particle might decay to a brane/particle with a different mass to charge ratio. Let us (following \cite{kn:motl}) interpret $m/q$ in (\ref{P}) as referring to the bulk physics state with the smallest possible mass/charge ratio, $\left(m/q\right)_{\textsf{min}},$ since such a state must be absolutely stable\footnote{An alternative proposal suggested in \cite{kn:motl} is that $m/q$ refers to the mass/charge ratio of the lightest charged brane/particle. The distinction will not matter in the present work, but ultimately it might: see below.}.

In discussions of the AdS/CFT correspondence, it is normally assumed that $\ell_5$ is by far the smallest length scale characterising the system, so we can take it that the left side of (\ref{P}) is ``small''. It cannot however be zero, for it is well known that the existence of charged brane/particles with arbitrarily small mass would lead to all manner of bizarre consequences: for example, the fine structure constant would run to zero at the macroscopic level. Certainly such brane/particles do not exist in our Universe or in the field theories arising on the boundary in the AdS/CFT correspondence, and duality implies that they do not exist in AdS physics either. Thus indeed the left side is strictly positive.

Turning now to the right side of (\ref{P}): the fact that it is not a constant (as of course its counterpart is in the asymptotically flat case, see the inequality (\ref{LL}) above) is not remarkable: the same holds true in the case of AdS$_5$-Reissner-Nordstr\"om black holes with spherical event horizons \cite{kn:102}. However, in that case, the analogous quantity on the right side is \emph{bounded below}, in fact by its limiting value for arbitrarily small charges: but, since the limit of small charges and masses is the same as the limit $L \rightarrow \infty$, this takes us back to the asymptotically flat case. That is, in both the asymptotically flat and the asymptotically AdS$_5$ \emph{spherically symmetric} cases, the inequalities analogous to (\ref{P}) can be interpreted as an \emph{upper bound} on $m\ell_5/q.$

The novelty here is that the right side of (\ref{P}) is \emph{not} bounded below as $Q^*$ becomes small; and if we try to take it to be arbitrarily small, then we immediately have a conflict with our discussion above. Thus we have to interpret this inequality in the opposite way: instead of interpreting it as an upper bound on the left side, we are forced to interpret it as a \emph{lower} bound on the right side. \emph{The Extremal Instability hypothesis forbids arbitrarily small charge parameters} for planar black holes.

In fact, as it stands, (\ref{P}) forbids the decay of any extremal black hole of this kind once its charge (and therefore mass) parameters become very small, that is, comparable to the left side (which we agreed is small, due to the presence of $\ell_5$). Following \cite{kn:kats, kn:NAH}, we ascribe this to the fact that (\ref{P}) is derived under the assumption that \emph{classical} Cosmic Censorship is valid. That is, when the charge and mass parameters of an extremal black hole of this kind fall to the point where the inequality is near to being saturated, we assume that quantum-gravitational effects modify the Censorship condition in such a manner that extremal black holes can continue to decay.

In our application, however, we can assume that if cold strongly coupled matter has a holographic dual, the black hole in the bulk is at least quasi-classical, so that classical Censorship continues to hold to a good approximation. We similarly assume (as is customary) that the decay rate is small enough that the black hole can be regarded as having approximately constant mass and charge (parameters) with respect to time. Under these circumstances, then, (\ref{P}) does impose a constraint on the charge parameter of planar black holes.

The principal conclusion of this Section is that the Extremal Instability hypothesis implies that, in the domain in which we are interested here, the charge parameter of the original extremal planar black hole cannot be prescribed arbitrarily. Let us explore the consequences.

\addtocounter{section}{1}
\section* {\large{\textsf{4. A Bound on the Density of Boundary Matter }}}
According to our discussion above, the left side of (\ref{P}) (with $m/q$ interpreted as $\left(m/q\right)_{\textsf{min}}$) is some strictly positive quantity, which is determined by the bulk particle spectrum and the bulk gravitational length scale. We now express the right side in a form suitable for a holographic interpretation.

We begin by discussing the meaning of the ``density'' of the strongly coupled matter we wish to study.

From equation (\ref{B}), we find that the electric potential in this case is
\begin{equation}\label{QUASAR}
\Phi\;=\;{k_5Q^*\over 2}\,\left({1\over r_{\textsf{H}}^2}\;-\;{1\over r^2}\right),
\end{equation}
where the constant term has been chosen, as usual, to ensure that the Euclidean potential one-form is well-defined throughout the Euclidean version of the spacetime. (See \cite{kn:clifford} for this procedure.) The limiting value of $\Phi$ as $r \rightarrow \infty$ yields \cite{kn:nat} the holographic (quark) chemical potential\footnote{Notice that, to construct a holographic model of cold quark matter, we need the black hole to be charged, for \emph{two} distinct reasons: first, so that we can drive down the Hawking temperature of the black hole and therefore the temperature of the boundary matter; and second, so that the large quark chemical potentials characteristic of cold quark matter can be represented.}, $\mu$, for the boundary matter:
\begin{equation}\label{QUACK}
\mu\;=\;{k_5Q^*\over 2 r_{\textsf{H}}^2}.
\end{equation}
From equation (\ref{I}), we see that, in the extremal case, this is just a certain multiple of $Q^{*{1\over 3}}.$

The enthalpy density of the boundary field theory, $\rho^{\textsf{enth}}$, is given holographically by the black hole mass per unit event horizon volume we computed above:
\begin{equation}\label{Q}
\rho^{\textsf{enth}} \;=\; {M^*\over r_{\textsf{H}}^3}.
\end{equation}
For the special case in which we are interested here, for which the temperature is approximately zero by the standards of strongly coupled quark matter, we need to evaluate this for near-extremal black holes. Substituting the right sides of (\ref{I}) and (\ref{L}) into (\ref{Q}), we have
\begin{equation}\label{R}
\rho^{\textsf{enth}}\,(T = 0)\; = \;{3\sqrt{3}\over 16\,\sqrt{2}\,\ell^{{5\over 2}}}\,\left({144\,k_5\,Q^{*2}\over \pi^5\,L^{10}}\right)^{{1\over 6}}.
\end{equation}
This too is a certain multiple of $Q^{*{1\over 3}},$ so we see that, in the holographic model of cold boundary matter, $\mu$ is a fixed multiple of $\rho^{\textsf{enth}}(T = 0)$:
\begin{equation}\label{RAT}
\mu \;=\; {4\sqrt{2\pi}\over 3\sqrt{3}}\,\sqrt{k_5}L\,\ell_5^{3/2}\,\rho^{\textsf{enth}}(T = 0).
\end{equation}
(Recall that $\mu$ has units of 1/(length), while of course the units of $\rho^{\textsf{enth}}(T = 0)$ are 1/(length)$^4$.)

This relation actually supplies the holographic interpretation of $k_5$: it is the parameter that controls the relation between the enthalpy density and the quark chemical potential of cold quark matter (given that the gravitational bulk parameters $L$ and $\ell_5$ are fixed: see below). That is, $k_5$ carries information about the field theory on the boundary.

For our purposes here, this simply means that we are justified in using either $\mu$ or $\rho^{\textsf{enth}}(T = 0)$ to measure the ``density'' of quark matter, and in fact they are used more or less interchangeably in the cold quark matter literature. Here we use $\rho^{\textsf{enth}}(T = 0)$.

Remarkably, we see that $\rho^{\textsf{enth}}\,(T = 0)$ is also a fixed multiple of the right side of (\ref{P}), and so (\ref{P}) itself can be written as
\begin{equation}\label{S}
\rho^{\textsf{enth}}\,(T = 0)\; > \;{3\,\sqrt{3}\over 4\,L\,\sqrt{2\pi\,k_5\,\ell_5^3}}\,\left({m\over q}\right)_{\textsf{min}}.
\end{equation}

We can use the ``holographic dictionary'' \cite{kn:nat} to eliminate $L$ (which is hard to interpret on the boundary): we have
\begin{equation}\label{T}
L\;=\;\left({2N_{\textsf{c}}^2\over \pi}\right)^{{1/3}}\,\ell_5,
\end{equation}
where $N_{\textsf{c}}$ is the number of colours characterising the boundary field theory. Thus we have
\begin{equation}\label{U}
\rho^{\textsf{enth}}\,(T = 0)\; > \;{3\,\sqrt{3}\over 4\,\left(32\,\pi\,N_{\textsf{c}}^4\right)^{{1\over 6}}\sqrt{k_5\,\ell_5^5}}\,\left({m\over q}\right)_{\textsf{min}}.
\end{equation}

The very \emph{existence} of such a bound, arising from considerations connected with the Weak Gravity Conjecture, is unexpected; and this is our main result. In view of the generality of the derivation, there is every reason to expect that, if one had a semi-realistic holographic description of QCD, then a bound of the same sort should exist in that case also. In other words, we have a direction in which to work towards a holographic account of the observed lower bound on the density of cold quark matter.

Actual applications of this result must await an explicit evaluation of the right side. We will not attempt that here, but we can offer a programme leading to that end.

The first point to make is the obvious one: in more complex holographic models of cold strongly coupled matter (such as those described in \cite{kn:nordita,kn:matti,kn:niko}), the inequality analogous to (\ref{U}) will surely look very different. Bearing that in mind, let us focus on the values of the parameters on the right, which will probably continue to be present in some form in such a generalization. The first step is to find the interpretations of the parameters in terms of the physics of the boundary field theory, so that the right side can (ultimately) be computed explicitly in terms of actual data.

We saw earlier that $k_5$ has a holographic interpretation in terms of the parameters of the field theory on the boundary, and of course this is also the case for $N_{\textsf{c}}$. It has recently been made explicit that the same is true of $\ell_5$. Let us explain.

Usually the holographic duality is applied to boundary systems for which gravitation is not important, such as the plasmas produced in heavy-ion collisions. However, cold quark matter exists, if at all, in systems like the interiors of very massive neutron stars, where gravitational fields are not negligible, so one needs a version of holographic duality in which gravitational fields can propagate on the boundary.

A concrete computational proposal for extending the duality in just this manner has recently been given in \cite{kn:boundary}. There, the initial conditions for the boundary state involve data on a spatial slice of the four-dimensional boundary, together with data on a bulk null slice meeting the specified boundary slice. The objective here is to use holography to gain a better understanding of the time evolution of the quantum energy-momentum tensor of strongly coupled matter in the presence of gravitation.

The two sets of given data must be consistent; enforcing this consistency is far from trivial \cite{kn:kroon,kn:horo}. Just as $k_5$ is a bulk parameter which is however partly determined by (strong interaction) boundary data, the consistency conditions entail a relation between $\ell_5$ and the four-dimensional Planck length $\ell_4$.

Finally, we need to consider $\left(m/q\right)_{\textsf{min}}$. Of course, $m$ has an interpretation in the boundary field theory in terms of the scaling dimension $\Delta$ (using formulae such as (\ref{LLLL}) above). As $\Delta$ is a monotonically increasing function of $m$, we need to compute a quantity of the form $\left(\Delta/q\right)_{\textsf{min}}$. Presumably this is possible in principle, though it is very far from clear how it can be done in practice. Of course, once again, even if it can be done, one can expect the result to be strongly modified in a more realistic model. Note that the question as to the precise interpretation of $\left(m/q\right)_{\textsf{min}}$, mentioned earlier, will be important here.

\addtocounter{section}{1}
\section* {\large{\textsf{5. Conclusion }}}
Observations of neutron stars suggest that even the enormous densities in their cores do not always suffice to generate cold quark matter \cite{kn:noquark}. There is a lower bound on the density of such matter, and the bound is very high.

We have seen that the holographic description of cold strongly coupled matter, in terms of an asymptotically AdS$_5$ black hole with flat transverse spatial sections and (near) zero Hawking temperature, can naturally explain the existence of a lower bound on the enthalpy density, if the Extremal Instability hypothesis is correct.

While we are not yet in a position to evaluate this bound explicitly, (\ref{U}) suggests that an ultimate holographic bound will be ``large''. This is indicated by the presence of the very small quantity $\ell_5^{5/2}$ which occurs on the right side of (\ref{U}) in its denominator. Thus, in the holographic picture, one might eventually be able to say that the lower bound on the density of cold quark matter is so high \emph{because gravity is so weak}. This brings us back full circle to the notion of ``weak'' gravity.

\addtocounter{section}{1}
\section*{\large{\textsf{Acknowledgement}}}
The author is grateful to Dr. Soon Wanmei for useful discussions.


\begin{thebibliography}{18}









\bibitem{kn:conf}
Roman Pasechnik, Michal \v{S}umbera, Different faces of confinement, Universe 7 (2021) 9, 330, arXiv:2109.07600 [hep-ph]
\bibitem{kn:urwitten}
Edward Witten, Anti-de Sitter Space, Thermal Phase Transition, And Confinement In Gauge Theories, Adv.Theor.Math.Phys.2:505-532,1998, arXiv:hep-th/9803131
\bibitem{kn:casa}
Jorge Casalderrey-Solana, Hong Liu, David Mateos, Krishna Rajagopal, Urs Achim Wiedemann, Gauge/String Duality, Hot QCD and Heavy Ion Collisions, in:
\emph{Gauge/String Duality, Hot QCD and Heavy Ion Collisions}, Cambridge University Press, Cambridge 2014, arXiv:1101.0618 [hep-th]
\bibitem{kn:nat}
Makoto Natsuume,	
AdS/CFT Duality User Guide, Lect.Notes Phys. 903 (2015), arXiv:1409.3575 [hep-th]
\bibitem{kn:bag}
Matteo Baggioli, A Practical Mini-Course on Applied Holography, SpringerBriefs in Physics 2019, arXiv:1908.02667 [hep-th]
\bibitem{kn:nickevans}
Nick Evans, Holography of Strongly Coupled Gauge Theories, EPJ Web Conf. 258 (2022) 08001, arXiv:2109.10121 [hep-ph]
\bibitem{kn:rocha}
Roldao da Rocha, Holographic entanglement entropy, deformed black branes and deconfinement in AdS/QCD, Phys. Rev. D 105 (2022) 026014, arXiv:2111.01244 [hep-th]
\bibitem{kn:noquark}
M.C. Miller, F.K. Lamb, A.J. Dittmann, S. Bogdanov, Z. Arzoumanian et al.,
The Radius of PSR J0740+6620 from NICER and XMM-Newton Data, Astrophys.J.Lett. 918 (2021) 2, L28, arXiv:2105.06979 [astro-ph.HE]
\bibitem{kn:robmann}
Chen Zhang, Robert B. Mann, Unified Interacting Quark Matter and its Astrophysical Implications, Phys. Rev. D 103, 063018 (2021), arXiv:2009.07182 [astro-ph.HE]
\bibitem{kn:nordita}
Carlos Hoyos, Niko Jokela, Matti J\"{a}rvinen, Javier G. Subils, Javier Tarrio, Aleksi Vuorinen, Holographic approach to transport in dense QCD matter, arXiv:2109.12122 [hep-th]
\bibitem{kn:matti}
Matti J\"{a}rvinen, Holographic modeling of nuclear matter and neutron stars, arXiv:2110.08281 [hep-ph]
\bibitem{kn:niko}
Niko Jokela, NICER view on holographic QCD, EPJ Web Conf. 258 (2022) 07004, arXiv:2111.07940 [hep-ph]
\bibitem{kn:sourya}
David Kastor, Sourya Ray, Jennie Traschen, Enthalpy and the Mechanics of AdS Black Holes, Class.Quant.Grav.26:195011,2009, arXiv:0904.2765 [hep-th]
\bibitem{kn:wilczek}
J. Preskill, P. Schwarz, A. D. Shapere, S. Trivedi, and F. Wilczek, Limitations on the
statistical description of black holes, Mod. Phys. Lett. A6 (1991) 2353–2362
\bibitem{kn:naresh}
Naresh Dadhich, K. Narayan, On the third law of black hole dynamics, Phys.Lett. A231 (1997) 335-338, arXiv:gr-qc/9704070
\bibitem{kn:haoong}
Hao Xu, Yen Chin Ong, Man-Hong Yung, Cosmic Censorship and the Evolution of d-Dimensional Charged Evaporating Black Holes, Phys. Rev. D 101, 064015 (2020), arXiv:1911.11990 [gr-qc]
\bibitem{kn:cygnus}
James C.A. Miller-Jones, Arash Bahramian, Jerome A. Orosz, Ilya Mandel, Lijun Gou et al., Cygnus X-1 contains a 21-solar mass black hole - Implications for massive star winds, Science 371 (2021) 6533, 1046-1049, arXiv:2102.09091 [astro-ph.HE]
\bibitem{kn:benson}
Andrew J. Benson, Arif Babul, Maximum spin of black holes driving jets, Monthly Notices of the Royal Astronomical Society 397 (2009) 1302 	
\bibitem{kn:motl}
Nima Arkani-Hamed, Lubos Motl, Alberto Nicolis, Cumrun Vafa, The String Landscape, Black Holes and Gravity as the Weakest Force, JHEP 0706:060,2007, arXiv:hep-th/0601001
\bibitem{kn:palti}
Eran Palti, The Swampland: Introduction and Review, Fortsch.Phys. 67 (2019) no.6, 1900037, arXiv:1903.06239 [hep-th]
\bibitem{kn:rude}
Ben Heidenreich, Matthew Reece, Tom Rudelius, Repulsive Forces and the Weak Gravity Conjecture, JHEP 1910 (2019) 055, arXiv:1906.02206 [hep-th]
\bibitem{kn:qing}
Qing-Guo Huang, Miao Li, Wei Song, Weak gravity conjecture in the asymptotical dS and AdS background, JHEP0610:059,2006, arXiv:hep-th/0603127
\bibitem{kn:naka1}
Yu Nakayama, Yasunori Nomura, Weak Gravity Conjecture in AdS/CFT, Phys. Rev. D 92, 126006 (2015), arXiv:1509.01647 [hep-th]
\bibitem{kn:mig}
Miguel Montero, A Holographic Derivation of the Weak Gravity Conjecture, JHEP 03 (2019) 157, arXiv:1812.03978 [hep-th]
\bibitem{kn:crem}
Sera Cremonini, Callum R. T. Jones, James T. Liu, Brian McPeak, Higher-Derivative Corrections to Entropy and the Weak Gravity Conjecture in Anti-de Sitter Space, JHEP 09 (2020) 003, arXiv:1912.11161 [hep-th]
\bibitem{kn:agar}
Prarit Agarwal, Jaewon Song, Large N Gauge Theories with Dense Spectrum and the Weak Gravity Conjecture, JHEP 05 (2021) 124, arXiv:1912.12881 [hep-th]
\bibitem{kn:naka2}
Yu Nakayama, Bootstrap bound on extremal Reissner-Nordstr\"om black hole in AdS, Phys.Lett.B 808 (2020) 135677, arXiv:2004.08069 [hep-th]
\bibitem{kn:102}
Brett McInnes, Holographic Dual of The Weak Gravity Conjecture, Nucl.Phys.B 961 (2020) 115270, arXiv:2007.05193 [gr-qc]
\bibitem{kn:103}
Brett McInnes, About Magnetic AdS Black Holes, JHEP 03 (2021) 068, arXiv:2011.07700 [gr-qc]
\bibitem{kn:104}
Brett McInnes, The Weak Gravity Conjecture Requires the Existence of Exotic AdS Black Holes, Nucl.Phys.B 971 (2021) 115525 arXiv:2104.07373 [gr-qc]
\bibitem{kn:105}
Brett McInnes, Extremal Bifurcations of Rotating AdS$_4$ Black Holes, JHEP 12 (2021) 155, arXiv:2108.05686 [gr-qc]
\bibitem{kn:heid}
Daniel Harlow, Ben Heidenreich, Matthew Reece, Tom Rudelius, The Weak Gravity Conjecture: A Review, arXiv:2201.08380 [hep-th]
\bibitem{kn:lemmo}
J.P.S. Lemos, Phys.Lett.B353:46,1995,
\x gr-qc/9404041; R.B. Mann, Class.Quant.Grav. 14 (1997) L109, arXiv:gr-qc/9607071;
Rong-Gen Cai, Yuan-Zhong Zhang, Phys.Rev.D54:4891,1996, \x gr-qc/9609065;
Danny Birmingham, Class.Quant.Grav. 16 (1999) 1197, arXiv:hep-th/9808032
\bibitem{kn:myers}
Robert C. Myers, Miguel F. Paulos, Aninda Sinha,
Holographic Hydrodynamics with a Chemical Potential, JHEP 0906:006,2009, arXiv:0903.2834
\bibitem{kn:peldan}
Dieter R. Brill, Jorma Louko, Peter Peldan,
Thermodynamics of (3+1)-dimensional black holes with toroidal or higher genus horizons,
Phys.Rev. D56 (1997) 3600, arXiv:gr-qc/9705012
\bibitem{kn:disc}
Weisstein, Eric W. "Polynomial Discriminant." From MathWorld--A Wolfram Web Resource. https://mathworld.wolfram.com/PolynomialDiscriminant.html
\bibitem{kn:surya}
Sumati Surya, Kristin Schleich, Donald M. Witt, Phase Transitions for Flat adS Black Holes, Phys.Rev.Lett.86:5231-5234,2001, arXiv:hep-th/0101134
\bibitem{kn:ruong}
Ru Ling, Hao Xu, Yen Chin Ong, How Anti-de Sitter Black Holes Reach Thermal Equilibrium, Phys.Lett.B 826 (2022) 136896, arXiv:2107.01556 [gr-qc]
\bibitem{kn:empa}
Tomas Andrade, Roberto Emparan, Aron Jansen, David Licht, Raimon Luna, Ryotaku Suzuki, Entropy production and entropic attractors in black hole fusion and fission, J. High Energ. Phys. (2020) 2020: 98, arXiv:2005.14498 [hep-th]
\bibitem{kn:pen}
Geoffrey Penington, Entanglement Wedge Reconstruction and the Information Paradox, JHEP 09 (2020) 002, arXiv:1905.08255 [hep-th]
\bibitem{kn:alm1}
Ahmed Almheiri, Netta Engelhardt, Donald Marolf, Henry Maxfield, The entropy of bulk quantum fields and the entanglement wedge of an evaporating black hole, JHEP 12 (2019) 063, arXiv: 1905.08762 [hep-th]
\bibitem{kn:alm2}
Ahmed Almheiri, Thomas Hartman, Juan Maldacena, Edgar Shaghoulian, Amirhossein Tajdini, Replica Wormholes and the Entropy of Hawking Radiation, JHEP 05 (2020) 013, arXiv: 1911.12333 [hep-th]
\bibitem{kn:seiberg}
Nathan Seiberg, Edward Witten, The D1/D5 System and Singular CFT,
JHEP 9904 (1999) 017, arXiv:hep-th/9903224
\bibitem{kn:niko1}
Oscar Henriksson, Carlos Hoyos, Niko Jokela, Novel color superconducting phases of N = 4 super Yang-Mills at strong coupling, JHEP 09 (2019) 088, arXiv:1907.01562 [hep-th]
\bibitem{kn:niko2}
Oscar Henriksson, Carlos Hoyos, Niko Jokela, Brane nucleation instabilities in non-AdS/non-CFT, JHEP 02 (2020) 007, arXiv:1910.06348 [hep-th]
\bibitem{kn:oscar}
Oscar Henriksson, Black brane evaporation through D-brane bubble nucleation, Phys. Rev. D 105, L041901 (2022), arXiv:2106.13254 [hep-th]
\bibitem{kn:107}
Brett McInnes, Planar Black Holes as a Route to Understanding the Weak Gravity Conjecture, arXiv:2201.01939 [gr-qc]
\bibitem{kn:maldacena}
Juan Maldacena, Liat Maoz, Wormholes in AdS, JHEP 0402 (2004) 053, arXiv:hep-th/0401024
\bibitem{kn:wit}
Edward Witten, Anti-de Sitter space and holography, Adv.Theor.Math.Phys. 2 (1998) 253, arXiv:hep-th/9802150 [hep-th]
\bibitem{kn:kats}
Yevgeny Kats, Lubos Motl, Megha Padi, Higher-order corrections to mass-charge relation of extremal black holes, JHEP 0712:068,2007, arXiv:hep-th/0606100
\bibitem{kn:NAH}
Nima Arkani-Hamed, Yu-tin Huang, Jin-Yu Liu, Grant N. Remmen, Causality, Unitarity, and the Weak Gravity Conjecture, arXiv:2109.13937 [hep-th]
\bibitem{kn:clifford}
Clifford V. Johnson, \emph{D-Branes}, Cambridge University Press, Cambridge, 2002
\bibitem{kn:boundary}
Christian Ecker, Wilke van der Schee, David Mateos, Jorge Casalderrey-Solana, Holographic Evolution with Dynamical Boundary Gravity, arXiv:2109.10355 [hep-th]
\bibitem{kn:kroon}
Diego A. Carranza, Juan A. Valiente Kroon, Construction of anti-de Sitter-like spacetimes using the metric conformal Einstein field equations: the vacuum case, Class. Quantum Grav. 35, 245006 (2018), arXiv:1807.04212 [gr-qc]
\bibitem{kn:horo}
Gary T. Horowitz, Diandian Wang, Gravitational Corner Conditions in Holography, JHEP 01 (2020) 155, arXiv:1909.11703 [hep-th]




\end{thebibliography}
\end{document}